\documentclass[aps,pre,floatfix,twocolumn,nofootinbib,showpacs]{revtex4} 
\usepackage{graphicx}
\newcommand{\BEQ}{\begin{equation}}
\newcommand{\EEQ}{\end{equation}}
\newcommand{\BEA}{\begin{eqnarray}}
\newcommand{\EEA}{\end{eqnarray}}

\renewcommand{\d}{{\rm d}}

\newcommand{\eps}{\varepsilon}
\newcommand{\half}{\frac{1}{2}}
\newcommand{\W}{{\cal W}}
\newcommand{\N}{{\cal N}}
\newcommand{\p}{p^{\downarrow}}
\newcommand{\pp}{\pi^{\downarrow}}
\newcommand{\maj}{\prec}
\begin{document} 
\draft
\title{Explanation of the Gibbs paradox within the framework of quantum thermodynamics}

\author{
A.E. Allahverdyan$^{1)}$ and Th.M. Nieuwenhuizen$^{2)}$}
\address{$^{1)}$Yerevan Physics Institute,
Alikhanian Brothers St. 2, Yerevan 375036, Armenia,\\
$^{2)}$ Institute for Theoretical Physics, 
University of Amsterdam,
Valckenierstraat 65, 1018 XE Amsterdam, The Netherlands 
}

\begin{abstract}

The issue of the Gibbs paradox is that when considering mixing of two
gases within classical thermodynamics, the entropy of mixing appears to
be a discontinuous function of the difference between the gases: it is
finite for whatever small difference, but vanishes for identical
gases. The resolution offered in the literature, with help of quantum
mixing entropy, was later shown to be unsatisfactory precisely where
it sought to resolve the paradox.
Macroscopic thermodynamics, classical or quantum, is unsuitable for
explaining the paradox, since it does not deal explicitly with the
difference between the gases.
The proper approach employs quantum thermodynamics, which deals with
finite quantum systems coupled to a large bath and a macroscopic work
source. 
Within quantum thermodynamics, entropy generally looses its dominant place and 
the target of the paradox is naturally shifted to the decrease of the maximally 
available work before and after mixing (mixing ergotropy). In contrast to entropy this is an
unambiguous quantity. For almost identical gases the mixing ergotropy
continuously goes to zero, thus resolving the paradox.  In this approach
the concept of ``difference between the gases'' gets a clear operational
meaning related to the possibilities of controlling the involved quantum
states.  Difficulties which prevent resolutions of the paradox in its entropic
formulation do not arise here.
The mixing ergotropy has several counter-intuitive features. It can increase
when less precise operations are allowed.  In the quantum situation (in
contrast to the classical one) the mixing ergotropy can also increase when
decreasing the degree of mixing between the gases, or when decreasing
their distinguishability. These points go against a direct association
of physical irreversibility with lack of information. 

\end{abstract}

\maketitle

\section{Introduction.}

Studying mixtures and mixing processes is one of the oldest tasks of
thermodynamics. Perhaps the most celebrated aspect of this task is the
Gibbs paradox: the entropy increase upon mixing two different gases 
stays finite for an arbitrary small difference between the gases, but is
zero for identical gases.

This paradox is discussed in many textbooks on
thermodynamics and statistical physics |e.g., in \cite{Landau,Balian}|
and it created a vast amount of literature during the last hundred
years till our days
\cite{G,S,T,Lande,Klein,LP,GLP,Lesk,VS,Kampen,Ca,DD,BG,Shu,Jaynes}.  
It was stated to be of a
high principal and methodical value \cite{S}, since it displays the
limits of applicability for classical (phenomenological)
thermodynamics: the resolution of the paradox (if any) ought to lie
 outside this discipline.  Already several times the paradox was claimed
to be resolved, but each time it was reconsidered and seen as an
open issue again.

The present status of the problem is somewhat controversial.  The
existing opinions can be roughly summarized as follows.

(1)\, The paradox is resolvable within the information theoretical
approach already in classical statistical physics \cite{VS,Ca}.
  
(2)\, The most natural resolution of the paradox has been achieved
within quantum statistical physics \cite{S,T,Lande,Klein,LP,GLP}
thanks to the feature of partial distinguishability.
  
(3)\, The quantum situation presents a natural setting for the
resolution, but there is a specifically quantum peculiarity of the
problem (induced by non-commutativity) which still prevents its
ultimate resolution \cite{DD}. Thus, the Gibbs paradox in quantum
statistical physics has so far not been resolved.

We share the last opinion.  Our purpose is to present an explanation of
this thermodynamical paradox starting from the first principles of
quantum mechanics.  

This is the program of {\it quantum thermodynamics}, see ~\cite{ABNQthermo} for a short review.
The crucial point in our discussion of the Gibbs paradox is to realize that it has
to be formulated in terms of the available work, as was already realized 
by Land\'e in 1955 \cite{Lande}.   (Within
the setup of classical thermodynamics this formulation is equivalent to
the entropic one \cite{Lande}).  In contrast to entropy, the available work 
-- by definition an ensemble average --
is a well-defined quantity for any equilibrium or non-equilibrium
state even of small quantum systems and it is a function of both the
state and the class of work sources employed for work extraction.
Moreover, the features of work are grounded directly on the first
principles of quantum mechanics.  On top of that, the amount of
available work adequately reflects intuition usually associated with
entropy, such as being a measure of non-equilibrium or disorder
\footnote{In fact the priority of the available work over entropy was
adequately understood already by Clausius; see discussions in
\cite{Corning}. Another example is Schr\"odinger, who in his famous book
\cite{What}, gave importance to (neg)entropy for characterizing survival
of organisms, but later on admitted that he should have spoken in this
context about the available work rather than entropy.  }. 

The above first-principles properties lead to the resolution of the
paradox in terms of mixing work which shows perfectly continuous
behavior when the difference between the gases goes to zero.
Difficulties which prevent resolutions of the paradox in its entropic
formulation do not arise here.  Certain aspects of the proposed scheme
|using work instead of entropy, time-scale separation, etc.| were
already anticipated in literature, e.g. in \cite{Lande,BG}. However,
these anticipations were conceived only in the framework of
phenomenological thermodynamics, and this is why the resolution in terms
of mixing work was not achieved \footnote{
The to be presented resolution of the Gibbs paradox resembles the recent
solution for the Maxwell demon problem presented by Scully and 
co-workers~\cite{ScullyDemon}: 
both find their basis in quantum thermodynamics, that is to say,
the thermodynamics of small quantum systems coupled to a macroscopic bath 
and work source, the latter leading to a time-dependent Hamiltonian.
Another recent result of quantum thermodynamics is 
our report on the breakdown of the Landauer inequality
for the energy needed to erase one bit of information ~\cite{Landauer}.}.

Our paper is organized as follows. In section \ref{classics} we recall
the classical formulation of the Gibbs paradox.  Next section reviews
the mixing entropy argument, an attempt to solve the paradox with help
of quantum entropy.  Section \ref{critique} discusses in detail why
this argument cannot be considered as a resolution of the paradox. Two
basic reasons for this are outlined and several pertinent issues
are discussed. Section \ref{main} presents the
resolution of the paradox with help of mixing work. In section
\ref{instruments} we discuss the mixing work in the contexts of
instruments available for work-extraction. The analysis fully embodies
the idea that the difference between two substances is first of all an
operational notion and should not be given any absolute status
\cite{Rosen}. Moreover, it appears that the dependence of the mixing
work on the available instruments is non-trivial, since it can both
increase or decrease upon introducing restrictions on those
instruments. Though the mixing work is zero when mixing identical
substances, in the quantum situation it can be a non-monotonous
function of the degree of mixing and of the (information-theoretic)
distinguishability between the mixing substances. These are shown in
sections \ref{degree} and \ref{dis}, respectively. The last section
presents our conclusions. Appendix \ref{takhtak} discusses definitions
of entropy and their relations to the second law; Appendix \ref{molod}
recollects several formulas.

\section{Classical formulation of the Gibbs paradox.}
\label{classics}

Consider two reservoirs each one having volume $V$.  They are
separated by a wall and are filled with different ideal
\footnote{For simplicity we choose to work with ideal gases. The
  ideality is not an issue for the Gibbs paradox: it exists for
  non-ideal gases as well \cite{GLP}, and the resolution obtained below for
  ideal gases will be generalizable to the non-ideal situation.}
Boltzmann gases, e.g., with two different isotopes of the same
substance. The difference is not specified, but assumed to be
tunable.  The number of particles $N$, pressure $P$ and temperature
$T$ in each reservoir are the same. The entropy $S$ of each gas is
\cite{Landau}
\footnote{\label{a1}In formula (\ref{entropy}) we omitted a term $Nf_m(T)$
with $f$ being some function of temperature, e.g., 
$f_m(T)=\frac{5}{2}-\frac{3}{2}\ln \left(  \frac{mT}{2\pi\hbar^2}\right)$ for a monoatomic gas.
This term does not
play any role in our discusion, since it drops out from the entropy difference. 
One also should not be troubled by the
presence of the dimension inside of the logarithm in (\ref{entropysimple}),
because it is canceled by the one of $f_m(T)$,
while in our further discussion it drops out anyhow
when calculating entropy differences.}
\BEA
\label{entropysimple}
S(N,V)=N\ln \frac{V}{N}
\EEA
Since the gases do not interact, the total entropy reads
\BEA
\label{2}
S_{\rm i} = S_1(N, V)+S_2(N, V)
=2N\ln \frac{V}{N}. 
\EEA

Now remove the wall. 
The overall system of the two gases is assumed to be thermally isolated
(the only influence of the external fields is in removing the wall)
\footnote{We shall focus on the mixing in the thermally isolated system.
For ideal gases this coincides with the isothermal mixing, since the
energy $U$ of such a gas depends only on its temperature:
$U/N=f_m(T)-Tf'_m(T)$, where the function $f_m(T)$ is discussed in Footnote
\ref{a1}. In general (i.e. for non-ideal gases), there will be a
difference between the isothermal mixing, where the temperature is kept
constant during the whole process with help of an external thermal bath,
and the mixing in the thermally isolated system, where the final temperature is
determined by the constancy of the overall energy. During the isothermal mixing
the gases will exchange some energy with the bath (mixing heat).
}.
The gases will mix, and after some transient time, a new equilibrium state is reached. 
Since in this state gases still do not interact, the final entropy $S_{\rm f}$ 
can be obtained again as a sum of two partial entropies, every component with $N$ particles
distributed in the volume $2V$,
\BEQ
S_{\rm f}= S_1(N, 2V)+S_2(N, 2V) =2N\ln \frac{2V}{N}
\EEQ
Thus the mixing entropy reads
\begin{equation}
\label{3}
\Delta S=S_{\rm f}-S_{\rm i} 
=2N\ln 2.
\end{equation}
The additional contribution $2N\ln 2$ arose due to the irreversible
process of mixing, and it {\it does not depend} on any quantitative
measure of the difference between the ideal gases. 

Now consider the same process, but assume that initially the gases are
identical. After removing the wall, Eq.~(\ref{3}) does not
predict any entropy change.  Indeed, in the final state we have a
one-component gas with total number of particles $2N$ in the volume $2V$. 
Thus, from Eq. (\ref{entropysimple}),
 $S_{\rm f}=2N\ln \frac{2V}{2N}$ and this equals $S_{\rm i}$, so $\Delta S=0$.
This is, of course, the expected and consistent
result, since there is no irreversibility when mixing two identical
gases in equilibrium; see in addition below and Footnote \ref{madras}. 
Thus we have arrived at the Gibbs paradox \cite{G,S,T}:
\footnote{\label{1}The paradox is not always formulated correctly in
  literature; see \cite{GLP} for detailed criticism. Some authors
  define entropy as $N\ln V$ and see the paradox in increasing the
  entropy when mixing two identical gases. Others think that the
  paradox is resolved by the very fact of not having any entropy
  increase when mixing identical gases.  To avoid confusion, we
  stress that the paradox is in the discontinuous change of entropy
  when tuning the difference between the gases. A closely related
  point|which can also be viewed as paradoxical|is that the mixing
  entropy does not depend on the actual difference between the gases,
  provided this difference is not zero. }

\begin{itemize}
  
\item When varying continuously the difference between the gases,
    the entropy defined according to Eqs.~(\ref{entropysimple},
\ref{2}, \ref{3}) changes
    {\it discontinuously}.

\end{itemize}

It is to be stressed that the existence of the Gibbs paradox is not
connected with the thermodynamic limit $N\to \infty$. As discussed in
\cite{LP,GLP}, the finite-$N$ situation does bring some differences in
the expression for the entropy of mixing, but the paradox survives; see
in addition Footnote \ref{madras}. 

\section{Mixing entropy argument.}
\label{argument}

\subsection{Assumptions of the argument.}
\label{pre}

It was realized by many scholars that the origin of the paradox is
that the difference between the gases is only assumed, but does not
show up explicitly in Eqs.~(\ref{entropysimple}--\ref{3}), i.e., the
description that led to the discontinuity is {\it not sufficiently complete}
\cite{Lande,Klein,LP,GLP}.  In that respect the paradox demonstrates
the limits of applicability of phenomenological thermodynamics. 

It is expected that for two ideal gases the difference will be related
to the internal states of their atoms \cite{Lande,Klein,LP,GLP,Lesk}:

(a)\,
Indeed, besides the translational motion which contributes to
  the entropy (\ref{entropy}), the atoms of the gases also have 
  internal states (e.g., spin states). These states are typically
  described by quantum mechanics.  For Boltzmann gases the
  internal states of the atoms are decoupled from the translational
  motion.  Returning to the above example of different gases in two
  reservoirs, let us assume that the first and second reservoirs
  contain atoms in internal states described by density matrices
\BEA
\label{vitsin}
\rho_1\qquad {\rm and}\qquad \rho_2,
\EEA
respectively \footnote{Recall that the density matrix|as well as the wave
function|refers to an ensemble of identically prepared
systems; see, e.g., \cite{Balian}.
Thus by ``state of a particle'' we necessarily 
mean the density matrix of the ensemble to which
this particle belongs. }. 

One of the main points in taking the internal states into account is
that now from the very beginning we can treat the two gases as
identical, but being in different internal states $\rho_1$ and
$\rho_2$ \cite{LP,GLP,Lesk}. This is similar to what happens in
nuclear physics, where the neutron and proton are considered as
identical particles (nucleons) in different states distinguished by
the value of the isotopic spin.
  
(b)\,
After removing the wall, the gases mix. We shall assume that the
  time-scale on which the internal states of the gases change is much
  larger than the time-scale related to mixing of the translational
  degrees of freedom.
  
(c)\,
Thus after the mixing, the internal states will be described by
  the density matrix ($M=2$)
\BEA 
\rho=\sum_{\alpha=1}^M \lambda_\alpha \rho_\alpha.
\label{mimi}
\EEA 
Since two equal amounts of gases are mixed,
the probability (weight) factors are equal,
$\lambda_1=\lambda_2=\half$. 

The same Eq.~(\ref{mimi}) applies for the mixing of $M$
gases with number of particles $\{N_\alpha\}_{\alpha=1}^M$ and
initial density matrices $\{\rho_\alpha\}_{\alpha=1}^M$; the
corresponding weights are
\BEA
\lambda_\alpha=\frac{N_\alpha}{\sum_{\alpha=1}^M N_\alpha},\qquad
\alpha=1,...,M.
\EEA
For the details of this generalization see section \ref{zanzibar}.

\subsection{Implementation of the argument \cite{Lande,Klein,LP,GLP}.}

Due to the above decoupling feature,
the total entropy of the translational motions and
the internal states of each gas is defined as [recall Footnote \ref{a1}]
\BEA
\label{entropy}
S_k(N,V)=&&N\ln \frac{V}{N}+N S_{\rm vN}(\rho_k),
\qquad k=1,2,
\\
S_{\rm vN}(\rho)&&\equiv-{\rm tr}\,\left[\rho\ln\rho\right],
\label{vN}
\EEA
where $\rho_k$ are given by (\ref{vitsin}), 
and where $S_{\rm vN}(\rho)$ is the von Neumann entropy.

The initial entropy of the two gases is the sum of two contributions
(recall that $N_1=N_2=N$)
\BEA
S_{\rm i}=2N\ln\frac{V}{N}+N S_{\rm vN}(\rho_1)+N S_{\rm vN}(\rho_2),
\EEA
while the final entropy reads
\BEA
S_{\rm f}
=2N\ln\frac{2V}{2N}+2N S_{\rm vN}(\rho).
\EEA
Recall that we treat two gases as identical; so in the final state
there is a single gas having $2N$ particles in volume $2V$.
The mixing entropy $\Delta S=S_{\rm f}-S_{\rm i}$ thus reads:
\BEA
\label{delta}
\Delta S
=2N\left[\,
S_{\rm vN}(\rho)
-\half S_{\rm vN}(\rho_1)-\half S_{\rm vN}(\rho_2)
\,\right],
\EEA

Assume that the internal states were maximally different, i.e., orthogonal,
\BEA
\label{serna}
\rho_1\rho_2=0.
\EEA
Such states can be distinguished by a single measurement, i.e., if it is
known that the state of a given single atom belongs to an ensemble
described by either $\rho_1$ or $\rho_2$, then a single measurement
suffices to establish the identity of the state.  In this respect
orthogonal states are similar to the classical case (perfect
distinguishability).  It is seen from definitions (\ref{mimi}, \ref{vN}) that
\BEA
S_{\rm vN}(\rho)=-{\rm tr}\left[
\frac{\rho_1}{2}\ln\frac{\rho_1}{2}
\right]-{\rm tr}\left[
\frac{\rho_2}{2}\ln\frac{\rho_2}{2}
\right],
\EEA
and that the mixing entropy $\Delta S=2N\ln 2$ 
agrees with the prediction (\ref{3}) of classical thermodynamics.

The other extreme is when the states are identical,
\BEA
\label{serna1}
\rho_1=\rho_2,
\EEA
which implies
$\Delta S=0$
again in agreement to the prediction of classical thermodynamics.

In general, if neither (\ref{serna}) nor (\ref{serna1}) is true, the
states $\rho_1$ and $\rho_2$ are only partially distinguishable,
i.e., any finite number of measurements will distinguish these states
with a finite error. Assume the states are pure:
\BEA
\rho_1=|a_1\rangle\langle a_1|\qquad {\rm and}\qquad
\rho_2=|a_2\rangle\langle a_2|.
\label{kaban}
\EEA
Noting the spectrum
\BEA
{\rm Spec}\,\left\{
\half |a_1\rangle\langle a_1|
+\half |a_2\rangle\langle a_2|
\right\} =
\half\left(
1\pm |\langle a_1|a_2\rangle|
\right),
\label{kabo}
\EEA
we get from (\ref{delta})
\BEA
&&\frac{\Delta S}{2N}=-{\rm tr}\left[\,\rho\ln \rho\,  \right] =h\left
(\frac{1-|\langle a_1|a_2\rangle|}{2}\right), \\
&&h(x)\equiv-x\ln x-(1-x)\ln (1-x).
\label{kaban2}
\EEA
This expression is minimal, and equal to zero for
identical gases $|\langle a_1|a_2\rangle|=1$.  It is maximal and equal 
to $2N\ln 2$ for totally distinguishable (orthogonal) states $|\langle
a_1|a_2\rangle|=0$.  In the intermediate case $0<|\langle
a_1|a_2\rangle|<1$, $\Delta S$ changes continuously, a conclusion that
holds more generally \cite{GLP}.
This was seen as a resolution of the Gibbs paradox 
\cite{Lande,Klein,LP,GLP,Lesk} \footnote{Note that there are several
  differences between the positions undertaken by the
  authors of \cite{Klein,LP,GLP} versus the one of Land\'e in
  \cite{Lande}.  The detailed analysis carried out in \cite{GLP}
  suggests that the approach by Land\'e contains errors, and his
  final formulas for the entropy of mixing are different from
  those in \cite{Klein,LP,GLP}.}. 
We shall recall counter arguments in section \ref{critique}.

\subsection{Generalization to several mixing gases.}
\label{zanzibar}

We shall indicate how (\ref{delta}) changes for the mixing of two gases
having initially non-equal number of particles and non-equal volumes.
The generalization to the mixing of several gases will be straightforward.

Let the first and second resevoirs contain,
respectively, $N_1$ and $N_2$ particles in volumes $V_1$ and $V_2$.
Since we are interested in irreversibilities coming due to mixing
only, we should assume that the initial pressures $P$ and temperatures
$T$ of the two gases are equal both initially and finally.  The known
ideal-gas relation $PV=NT$, applied for $V=V_1,\, V_2,\, V_1+V_2$ and
$N=N_1,\, N_2,\, N_1+N_2$, implies \BEA
\label{es}
\frac{P}{T}=
\frac{N_1}{V_1}=\frac{N_2}{V_2}=\frac{N_1+N_2}{V_1+V_2}.
\EEA

Using (\ref{es}) and 
proceeding along the same lines as when deriving (\ref{delta}),
we get for the mixing entropy ($M=2$)
\BEA \label{est}
\frac{\Delta S}{\sum_{\gamma=1}^M N_\gamma}
=S_{\rm vN}\left(\sum_{\alpha=1}^M\lambda_\alpha\rho_\alpha\right
)-\sum_{\alpha =1}^M\lambda_\alpha S_{\rm vN}(\rho_\alpha),
\EEA
where 
\BEA
\label{kov}
\lambda_\alpha=\frac{N_\alpha}{\sum_{\gamma=1}^M N_\gamma}, \qquad \alpha=1,..,M.
\EEA
are the fractions of the two gases in the final density matrix.
We already wrote Eqs. (\ref{est}--\ref{kov}) such that they hold for any $M\ge 2$.

\section{Critique of the quantum mixing entropy argument.}
\label{critique}

\subsection{Thermodynamic entropy of mixing is 
ill-defined in quantum mechanics.}

The above argument on the continuous change of $\Delta S$ was seen by
many as the resolution of the Gibbs paradox
-- and it is often still believed to be. However, a more detailed
analysis has shown that this explanation creates a new conceptual problem \cite{DD}.
Let us recall the following features of the thermodynamical entropy:
\begin{itemize}
  
\item If two states A and B are connected by an irreversible process
  ${\rm A}\to {\rm B}$, then for defining thermodynamically the
  entropy change during this process, we should connect those states
  by a certain reversible process ${\rm A}\Rightarrow{\rm B}$ |possibly 
by involving thermal baths and sources of work|and
  calculate the entropy change $\Delta S$ via the Clausius formula 
\BEA
\label{cl}
\Delta S ({\rm A}\to{\rm B})=\int_{{\rm A}\Rightarrow{\rm B}}
\frac{\d Q}{T},
\EEA
where $\d Q$ and $T$ are, respectively, the differential heat 
(received from thermal baths) and the temperature.

Eq.~(\ref{cl}) provides entropy with an operational meaning and makes it
observable via macroscopic measurements. Indeed, determining, e.g., the
von Neumann entropy via its definition (\ref{vN}) implies knowledge of
{\it all} eigenvalues of the corresponding density matrix $\rho$. This
knowledge is not available for typical macroscopic or mesoscopic systems. 

\item A reversible process ${\rm A}\Rightarrow{\rm B}$ is defined by
  requiring that it is possible to pass back along the same trajectory
and to return to the same thermodynamical state 
\footnote{\label{hawk}Thermodynamical state is defined by the values of certain macroscopic
quantities, such as pressure, temperature, magnetization, entropy, etc.}, 
such that, in particular, the entropy change during the resulting 
cyclic process 
${\rm A}\Rightarrow{\rm B}\Rightarrow{\rm A}$ is equal to zero:
$\int_{{\rm A}\Rightarrow{\rm B}\Rightarrow{\rm A}}
\frac{\d Q}{T}=0$.

\end{itemize}

Any statistical definition of entropy is expected to agree with the above
thermodynamical one. An inspection shows, however, that this is not
the case \cite{DD}: the partially distinguishable (i.e.,
non-orthogonal) states|which were supposed to solve the paradox | create 
in this respect an inconsistency.  
It appears that for such states there is no reversible mixing process.
Let us first of all note
that when the internal states $\rho_1$ and $\rho_2$ are orthogonal | that 
is, they correspond to definite eigenvalues $a_1$ and $a_2$ of
some physical observable (hermitean operator) $A$|it is possible to
separate the mixed gases, and at least in principle to fulfil the
requirement of a cyclic process.
What one needs for this purpose is a suitable Hamiltonian
\cite{Peres}
\BEA
\label{bp}
H_{\rm sep}
=f(\vec{r}, A),
\EEA
which establishes strong correlations between the internal states of the
atoms and their translational motion described by the position vector
$\vec{r}$ \cite{Peres}: the function $f(\vec{r}, a_{i})$, with $i=1,2$,  
is very small for $\vec{r}$ being in, respectively, first and second reservoirs.
The magnitude of $H_{\rm sep}$ has to be sufficiently large, so that all 
other terms in the overall Hamiltonian can be neglected.
Together with a low temperature bath, weakly coupled to the gases, the
Hamiltonian $H_{\rm sep}$ will drive the system towards its minima and it 
will separate the mixed gases back into different
reservoirs \cite{Peres} without changing the internal states of the atoms
(since $[H_{\rm sep},A]=0$).
There can be practical limitations on this procedure
related, e.g., with limitations on the magnitude of $H_{\rm sep}$, but
in principle such a process is possible. Thus, one can apply 
(\ref{cl}) and recover of the usual thermodynamical formulas for
entropy \cite{Peres}. 

The problem is that once the gases described by initially {\it partially}
distinguishable (non-orthogonal) density matrices $\rho_1$ and $\rho_2$
(state A) are mixed with weights $\lambda_1$ and $\lambda_2$,
respectively (state B), then it is impossible to go back to the original
state by any process such that the two gases return to their original
states: There is {\it no} Hamiltonian similar to $H_{\rm sep}$ in (\ref{bp})
which can achieve such a separation \cite{Peres}, in particular because
$\rho_1$ and $\rho_2$ do not form eigenstates of any hermitian operator. 

Are there, however, measurements which can help to achieve this separation?
We need a careful discussion of this question, since the existing opinions|e.g., 
those presented in \cite{DD}|seem to us somewhat unclear.

First of all, we note that the procedure involving $H_{\rm sep}$ can be
seen as a measurement, where the role of the measuring apparatus is
played by the classical coordinate $\vec{r}$ of the atom 
\footnote{A closely related quantum mechanical model for quantum (and classical) 
measurements was  recently analyzed in detail in collaboration
with R. Balian, Ref.~\cite{ABNSwedish}}. 
The motion of this apparatus amounts to the separation of the gases. 
The above question can be thus reformulated as to concern other measuring 
apparatuses (not connected with the coordinates) and their role for separation of
the gases. Our answer to this question is negative, and here is why.

Using the example given by (\ref{kaban}), it is seen that there is not
any measurement which would discriminate unambiguously|and without
disturbing the initial states|between $|a_1\rangle$ and $|a_2\rangle$,
if $\langle a_2|a_1\rangle$ is neither zero nor one \cite{Peres}.
Thus, it is impossible to separate the gases without disturbing the
states of their atoms. However, requiring cyclic changes of every
single atom state is too much for a thermodynamical reasoning. It suffices
to require cyclic change of all collective (macroscopic variables) of
the gases.  In particular, the (final) internal states of the atoms in each
reservoir are to be described by the density matrices
$\rho_1=|a_1\rangle\langle a_1|$ and $\rho_2=|a_2\rangle\langle a_2|$,
respectively.  Such (generalized) measurements do exist \footnote{This
  is a known fact in the physics of quantum ensembles; see, e.g.,
  Ref.\cite{Erwi}.  The described procedure amounts to POVM (positive
  operator measured values).  Recently we discussed in detail its
  implications for defining fluctuations of work \cite{ANbkj}.}.
Assume for simplicity that the internal state is a spin-$\half$
represented by Pauli matrices
$\vec{\sigma}=(\sigma_1,\,\sigma_2,\,\sigma_3)$.  One comes with
another set of particles carrying spin-$\half$ described by Pauli
matrices $\vec{s}=(s_1,\,s_2,\,s_3)$.  The spins $\vec{\sigma}$ and
$\vec{s}$ undergo a controlled unitary evolution, after which one
measures, e.g., $s_3$ with help of a suitable macroscopic measurement
apparatus.  After selecting measurement results (i.e., the eigenvalues
$\pm 1$ of $s_3$), the 
initial mixed ensemble
$\rho=\lambda_1|a_1\rangle\langle a_1|+\lambda_2|a_2\rangle\langle a_2|$
of the $\vec{\sigma}$ spin is separated
into two subensembles
$\rho_1=|a_1\rangle\langle a_1|$ and $\rho_2=|a_2\rangle\langle
a_2|$, with the probabilities (weights) $\lambda_1$ and $\lambda_2$,
respectively
\cite{Erwi,ANbkj}.  
This is the desired separation.

However, quantum measurements are by their very nature {\it
  non-cyclic}, since dissipative processes are connected with the motion of the
pointer variable. In the above example both the spin $\vec{s}$ and the
apparatus measuring $s_3$ have undergone such non-cyclic processes.  These
certainly do generate an independent (and sizable) amount of entropy
which is not taken into account in (\ref{cl}).

\begin{itemize}
\item In summary, the possibility to define a cyclic process is a
  necessary condition for the thermodynamical meaning of entropy.
  When mixing gases that have non-orthogonal states, there are no suitable
  cyclic unmixing processes. This precludes entropy from having
  the proper thermodynamical meaning.  Thus, trying to solve the
  problem in one place the quantum mixing entropy argument creates a
  new problem almost at the same time. The Gibbs paradox thus remains
  unexplained.

\end{itemize}

\subsubsection{Why it is impossible simply to define entropy via the von
Neumann formula?}
\label{contraentropy}

In the context of the above objection to the thermodynamical meaning of the mixing
entropy in the quantum situation, one can ask why it is not possible
simply to {\it define} entropy via the von Neumann formula (\ref{vN})
without worrying on its precise relation to other thermodynamical
notions.  If desired, such a definition may be motivated, e.g., via
information-theoretic arguments \cite{Balian}. 

In our opinion this is not possible to do, since entropy in
statistical physics is never defined as an independent macroscopic
observable; note again that the calculation of the von Neumann entropy
via Eq.~(\ref{vN}) requires the knowledge of the full spectrum of the
density matrix $\rho$, which is microscopic information normally not
available for statistical systems. 
For internal states, it may be available, though.
 More generally, entropy cannot be
defined from first principles without taking into account the
corresponding formulations of the second law of thermodynamics, a fact
that strictly speaking precludes any really non-circular derivation of
these entropic formulations from first principles \cite{Landau}.
In contrast, formulations of the second law that operate with work
instead of entropy normally do have first principle derivations; see,
e.g., \cite{Lenard,Lindblad,Minima}.  The non-unique character of
entropy is recalled and illustrated in Appendix \ref{takhtak}.

\subsection{The employed notion of ``difference between gases'' 
  does not have a clear operational status.}
\label{o}

Another difficulty with the above argument is that this attempted  
resolution of
the paradox does not depend on the available experimental instruments
and tools to be employed in control of the internal states of the
atoms.  As it stands within the entropic argument, the resolution
depends on the difference between the states which is determined by
their initial preparations via density matrices $\rho_1$ and $\rho_2$.
However, preparation and control are different things and in general
cannot be combined in a single density matrix.  As an example, consider
preparation of a Gibbsian state with density matrix $\rho\propto
\exp[-H/T]$, where $H$ and $T$ are, respectively, the Hamiltonian and
the temperature. This preparation needs only a weak
interaction between the system and a thermal bath at temperature $T$;
it does not contain any information on what we can measure or control
in this system.

On general grounds, it was argued in \cite{Rosen} that the resolution of
the Gibbs paradox {\it has to be} operational, since there are
situations when two objects are formally different, but no computable
(i.e., solvable by algorithms) operation can establish this difference.
Worse, we cannot exclude unknown laws of physics that in the future
would force us to distinguish (states of) atoms or particles which in
our present understanding are considered as identical.

This operational aspect is also important, because, as we see below for
the approach that takes this properly into account, the
dependence on the available instruments is non-trivial: less refined
instruments can|depending on the situation|indicate less or more
irreversibility of mixing.

One may perhaps counterargue the above criticism by noting that the
operational meaning and the dependence on the available instruments
might be provided by the information-theoretic approach to
statistical physics; see, e.g., \cite{VS,Ca}. We, however, should
simply note that information-theoretic constructions are not at all
guaranteed to have the proper physical meaning, as we saw for the
above reversibility problem. Moreover, uncritical use of
information-theoretic concepts may by itself lead to problems; see, e.g.,
\cite{Landauer}, where the first-principle derivation of the Landauer
bound for information erasure was found in conflict with the
information-theoretical one.

\section{Resolution of the paradox via 
the concept of maximal mixing work.}
\label{main}

The main point of the present paper is to employ quantum thermodynamics
|the thermodynamics of finite systems coupled to a macroscopic worksource
and possibly to a macroscopic bath. As realized in earlier works
~\cite{Landauer,NA,ABNQthermo},
this approach generally acknowledges that one should study work
instead of entropy -- in the absence of a thermodynamic limit the latter
has no firm meaning and each definition leads to a new value.  This
shift of paradigm will allow us to resolve the Gibbs paradox without the
difficulties and ill-defined meaning of the mixing entropy argument.
The reason for this solution lies in the fact that work and its
properties are deduced from the first principles of quantum mechanics
without any need of thermodynamic postulates (such as reversibility or
existence of cyclic processes); see in this context Footnotes
\ref{conrad1} and \ref{conrad2}. In other words, the resolution of the
paradox is sought by going to the first principles of quantum mechanics
alone, and {\it without involving any thermodynamic argument}. 

We start by recalling the definition of available work for a general,
thermally isolated process done on a quantum system.

\subsection{Definition of work.}
\label{wowork}

A quantum system is described at the initial time $t=0$ by
a density matrix $\rho(0)$ and
interacts with an external macroscopic work source.
The resulting evolution of the system is generated by (an effective)
Hamiltonian $H(t)=H\{R(t)\}$, which is time-dependent via classical
(c-number) parameters $R(t)$ (control fields).

We shall be concerned with processes where the change of the Hamiltonian is cyclic
\BEA
\label{cycle}
H(\tau)=H(0)=H.
\EEA
The situation where the work-source interacts with the system for a
finite time belongs to this class of processes, since the
corresponding system-work-source interaction Hamiltonian is zero both initially
and finally \footnote{All constructions below generalize to processes,
where the initial and the final Hamiltonians are different.
In the context of the Gibbs paradox this more general setting may
provide some advantages, though it does not give any conceptual novelty
as compared to the cyclic-Hamiltonian case.}.
Note that processes with a cyclic Hamiltonian are obviously different from the processes
that are cyclic in the sense of various macroscopic quantities. However, 
it is necessary to have a cyclic change of the Hamiltonian for the process to be 
cyclic in the sense of macroscopic quantities.

Thus, the process is assumed to be thermally isolated and 
the Hamiltonian $H(t)$ generates a unitary evolution:
\BEA 
\label{evolution} 
&&i\hbar\frac{\d}{\d t}{\rho}(t)=[\,H(t),\rho(t)\,],\\
\label{ddd}
&&\rho(t)=U_t\,\rho(0)\,U_t^\dagger, ~~~
U_t=\overleftarrow{\exp}\left[-\frac{i}{\hbar}\int_{0}^{t}\d s\,
H(s)\right],~~~~
\label{unita}
\EEA 
where $\overleftarrow{\exp}$ denotes the time-ordered exponent.  It is
well known that, in general, a Hamiltonian evolution for two coupled
systems does not reduce to a Hamiltonian evolution for one of them.
However, in the present case the evolution of the
system is Hamiltonian owing, in particular, 
to the {\it macroscopic} character of the
work-source, as discussed in \cite{Balian, Dom_power}
\footnote{The appendix of Ref.~\cite{Dom_power} contains a clear
discussion of certain additional conditions that have to be satisfied
for the time-dependent Hamiltonian evolution and for the proper
identification of the work-source.  }.

\begin{itemize}
\item The work $W$ done by the external source between times $0$ and $\tau$
in the thermally isolated process
\footnote{\label{conrad1}From the viewpoint of work-exchange every
process can be completed to a thermally isolated one by including in the
system its environment (e.g., thermal baths). Then the work (\ref{work})
for this thermally isolated process coincides with the usual definition
of work for an arbitrary process: $W=\int_0^\tau\d t\,{\rm tr}\left(
\rho_S(t)\,\partial_t H(t) \right)$, where $\rho_S(t)$ is
the time-dependent density matrix of the system.  Indeed, let $H_{E}$
and $H_{I}$ be, respectively, the Hamiltonian of the environment and the
system-environment interaction. Recall that the work-sources act only on
the system; thus the total Hamiltonian ${\cal H}(t)$ of the system+environment 
is ${\cal H}(t)=H(t)+H_E+H_{I}$, where only the system Hamiltonian $H(t)$ 
is time-dependent. To prove the desired statement we have to
write down the expression (\ref{work}): $W={\rm tr}[\rho_{SE}(\tau)\,
{\cal H}(\tau)- \rho_{SE}(0)\,{\cal H}(0)]$, where $\rho_{SE}(t)$ is the time-dependent
density matrix of the system+environment, apply the von 
Neumann equation of motion for the thermally isolated process:
$i\dot{\rho}_{SE}=[{\cal H}, \rho_{SE}]$, and transform 
$W=\int_0^\tau\d t\,{\rm tr}\left(\rho_S(t)\partial_t H(t) \right)
=\int_0^\tau\d t\,{\rm tr}\left(
\rho_{SE}(t)\partial_t {\cal H}(t)\right)$ with help of integration by parts.
}
is identified with the average energy change of the system \cite{Landau,Balian}
\BEA 
W={\rm tr}[\rho(\tau)\,H(\tau)- \rho(0)\,H(0)]
\label{work} 
\EEA 

\item Due to conservation of [average] energy, $W$ is equal to the
  average energy decrease of the work source. 

\item This is a classical, mechanical energy that can be transferred with
100\% efficiency to other macroscopic work-sources, and, in
particular, it can transferred to another mechanical degree of freedom
performing classical deterministic motion.
  
\item $W$ is typically observed via suitable (classical) measurements
  done on the macroscopic work source, or, alternatively, by measuring the initial
  and final average energies on the ensemble of (many) identically prepared
  systems.  Both these ways are routinely employed in practice, e.g.,
  in NMR/ESR physics, where the system corresponds to
  spin-$\frac{1}{2}$ under influence of external magnetic fields
  (work-source).
  
\item 
The definition of work and its features are based purely on the first
principles of quantum mechanics. They do not depend on thermodynamical
concepts, such as reversibility 
\footnote{\label{conrad2}In particular,
the definition of a reversible process can be based on the notion of
work \cite{Perrot}.  A process is reversible if {\it i)} it can be
supplemented by its mirror reflection that goes back along the same
trajectory; {\it ii)} the work done on this completed process is zero.
It is also clear that the definition of heat need not supersede the
definition of work. The reason for this is that from the viewpoint of
work-exchange any process can be completed to a thermally isolated one,
where the work is uniquely related to the energy; see Footnote
\ref{conrad1} in this respect. }.  In contrast, the work as it is known in
thermodynamics can be deduced from the first principles of quantum
mechanics.

\end{itemize}

\subsection{Maximally available work.}
\label{maxi}

One of the fundamental tasks of thermodynamics is to determine the
maximal amount of work which can be extracted from a given
(non-equilibrium) system in the initial state $\rho$ under cyclic-Hamiltonian
(sufficiently smooth) processes (\ref{cycle}).
The latter
condition is imposed, since otherwise there may not be any limit in
the extracted work (e.g., for the final Hamiltonian being negative and very
large by the absolute value). It is via this task posed by Clausius and solved
within phenomenological thermodynamics that entropy acquires
its physical meaning as a measure of order related to high-graded
energy (work) \cite{Landau,ABN}.  While the standard solution of this
task is well known and based on the notion of reversible process
(in the same way as the definition of entropy is), it was recently
shown that the problem can, and should, be solved from the first principles of
quantum mechanics {\it without} invoking any thermodynamical axiom
\cite{ABN,Dom_power}. 
The solution differs from the standard one, though the latter provides
a correct bound for the maximal work $W_{\rm max}$ in (\ref{nana})
below, and is expected to agree with it when phenomenological
thermodynamics is supposed to apply, i.e., for weakly non-equilibrium
states of generic macroscopic systems. 

To describe the solution to the
maximal work extraction problem, we denote the eigenresolutions of the
Hamiltonian $H$ and of the density matrix $\rho$ as, respectively,
\BEA
\label{resol}
H=\sum_{k=1}^n
\eps_k|\eps_k\rangle\langle\eps_k|, \qquad
\rho=\sum_{k=1}^n
p_k|p_k\rangle\langle p_k|,
\EEA
where $\{|\eps_k\rangle\}_{k=1}^n$ and $\{|p_k\rangle\}_{k=1}^n$ with
$\langle\eps_k|\eps_l\rangle=\langle p_k|p_l\rangle =\delta_{kl}$ are
the eigenvectors of $H$ and $\rho$, respectively, and where $\eps_k$
and $p_k$ are the corresponding eigenvalues. We shall assume that 
always be ordered as
\BEA
\label{hu}
\eps_1\leq\eps_2\leq....
\EEA
The non-increasing ordering of $\{|p_k\rangle\}_{k=1}^n$ is denoted as 
\BEA
\label{shut}
\p_1\geq \p_2\geq....
\EEA
Then the maximal available work is defined as \cite{ABN}
\BEA
\label{nana}
W_{\rm max}\equiv -\W=
{\rm min}_{\,U}
\{\,
{\rm tr}\,H[\rho(\tau)- \rho(0)]\},
\EEA
where $\W$ is the non-negative absolute value of the maximal work,
and where the minimization in ${\rm min}_{\,U}$ is taken over 
{\it all} smooth, cyclic 
Hamiltonians
\footnote{\label{gell}Note that for an $n$-level system the
  minimization over all Hamiltonians (\ref{kura}) can be carried out
  by minimizing over Hamiltonians of the form
  $H(t)=H+\sum_{i=1}^mb_i(t)X_i$, where $b_i(t)$ are time-dependent
  c-functions, and where $X_i$ are operators such that any generator
  of the group SU(n) can be obtained via linear combinations of
  $H,\,X_1,\,X_2,...,X_m$ and their multiple commutators \cite{samo}.
  For $n=2$ and $H=\sigma_3$ this Hamiltonian is
  $H(t)=\sigma_3+b(t)\sigma_1$, with $\sigma_1$ and $\sigma_3$ being
  the corresponding Pauli matrices. For $n=3$ the analogous
  Hamiltonian is $H(t)=\lambda_3+b_1(t)\lambda_1+b_4(t)\lambda_4$,
  where $\lambda_k$ are the Gell-Mann matrices [generators of SU(3)].\\
  If the minimization is carried out via Hamiltonians
  $\lambda_3+b_1(t)\lambda_1$, the unitary transformations act only on
  the upper left $2\times 2$ sector of the $3\times 3$ density matrix
  $\rho$.}
\BEA
\label{kura}
H(t)=H+V(t),\qquad V(0)=V(\tau)=0,
\EEA
where $\tau$ is the cycle time of the Hamiltonian
\footnote{\label{mell}We note that there are no
  restrictions on the product of $\tau$ with the typical magnitude of
  $V(t)$ (i.e., on the dimensionless coupling constant characterizing  the sources of
  work). It is also assumed that the initial state $\rho$ is
  known. Limitations on this knowledge will, in general, lower the value
  of the maximal work.}. 
Minimizing over the Hamiltonians in (\ref{kura})
is equivalent to minimizing over all unitary operators $U$
\cite{ABN}; this is why we denoted this minimization as ${\rm min}_{\,U}$. 
An explicit formula for the optimal Hamiltonian is given
in \cite{ABN}, while the result of the minimization in (\ref{nana})
yields the {\it ergotropy} \cite{ABN,Dom_power}:
\BEA
\W={\rm tr}\,\left(\rho H\right)-\sum_{k=1}^n\p_k\eps_k\geq 0.
\label{na}
\EEA
This is a difference between the final and initial average energies of
the system, as it should be for the work extracted in a thermally
isolated system. It has a simple interpretation: since in quantum
mechanics the eigenvalues of $\rho$ are conserved under the unitary
evolution caused by macroscopic external sources, the lowest final
energy is reached when the largest eigenvalue of $\rho$ becomes the
ground-state occupation, the one but largest eigenvalue occupies the
first excited state and so on.  Various features of $\W$, in particular
those contrasting the thermodynamical intuition, were studied in
\cite{ABN,Dom_power}.  We suggested to call (\ref{na}) the {\it ergotropy}
of the state $\rho$. 

It is seen from (\ref{na}) that no work extraction is possible (i.e.,
$\W=0$) if $\rho$ is a monotonically decreasing function of $H$:
\BEA
\label{thomson}
\rho=f(H),\qquad f'(x)\leq 0.
\EEA
This, in particular, includes Gibbs equilibrium states $\rho\propto
e^{-H/T}$, where $T>0$ is the temperature. This confirms 
Thomson's formulation of the second law: no work extraction 
from an equilibrium state by means of cyclic-Hamiltonian processes
\cite{Lenard,Lindblad}.

\subsection{The operational meaning of the available work.}
\label{oper}

The concept of maximal work takes into account the notion of available
instruments. Indeed, in (\ref{nana}) we optimized the extracted work
over all cyclic-Hamiltonian thermally isolated processes, which assumes that the
optimal one is available. If there are restrictions on the
availability of sources of work, the amount of extractable work will,
in general, be smaller than $\W$. It is even possible that no work at
all can be extracted by some restricted class of work sources 
\footnote{
The class of employed work-sources corresponds to
what in \cite{BG} was called a thermodynamical construction: a set of
non-relaxed mechanical degrees of freedom that define the very meaning
of various thermodynamical quantities.}.

To make this point clear, let us
assume that the possible unitary evolutions $U$
in (\ref{nana}) are restricted to permutations of the diagonal
elements 
\BEA
\pi_k=\langle \eps_k|\rho|\eps_k\rangle
\EEA
of the density matrix $\rho$ in the energy
representation 
\footnote{There are, of course, many other ways to introduce
  limitations on the available unitary evolutions. 
  For more examples, see Ref.~\cite{Dom_frame}, as well as, the last
  part of Footnote \ref{gell} and Footnote \ref{mell}.}.
  Then instead of (\ref{na}) we will have
\BEA
\label{klm}
W'_{\rm max}\equiv -\W'(\rho,H)&&=
\sum_{k=1}^n\pp_k\eps_k-{\rm tr}\,\left(\rho H\right)\\
&&=\sum_{k=1}^n
\eps_k\left[\pp_k-\pi_k\right]
\EEA
In general, we have for the ergotropy 
\BEA
\label{klm1}
\W\geq\W',
\EEA
where the equality sign is realized for $[\rho,H]=0$.
It is now possible that $\pp_k=\pi_k$ and thus $\W'=0$, though $\W>0$
due to the non-diagonal elements of $\rho$.

\subsection{Explanation of the paradox.}

We shall now immediately deal with $M$ gases with arbitrary
weights $\lambda_\alpha=\frac{N_\alpha}{\N}$, and the
total of particles
\BEA
\N=\sum_{\alpha=1}^M N_\alpha.
\EEA
Let us return to the assumptions presented in section \ref{pre} and list
them again:
{\it i)} The necessity of taking into account the internal states.
{\it ii)} Decoupling of the internal and translational degrees of freedom:
the total Hamiltonian $H^{(\alpha)}_{\rm tot}$ of each gas contained 
in the corresponding reservoir is 
\BEA 
H^{(\alpha)}_{\rm tot}=H^{(\alpha)}_0+\sum_{i=1}^{N_\alpha} H^{(\alpha,i)},
\quad \alpha=1,...,M.
\EEA
where $H_0$ is the sum of kinetic energies of all $N$ gas particles
plus the potential generated by the walls of the reservoir, and where
$H^{(\alpha,i)}$ is the Hamiltonian of internal motions of the
atom with index $i$ belonging to the gas with index $\alpha$. 
Since we assume that all atoms in both reservoirs are
identical and differ by their states only, we shall assume that all
atoms have the same internal Hamiltonian:
\BEA
H^{(\alpha,i)}=H.
\EEA

{\it iii)}
Time-scale separation between the translational and internal degrees
of freedom during the mixing; thus Eq.~(\ref{mimi}), 
$\rho=\sum_{\alpha=1}^M\,\lambda_\alpha\rho_\alpha$,
 holds for the post-mixed
density-matrix for $M$ gases with the initial internal states $\rho_\alpha$ and
arbitrary weights $\lambda_\alpha$.

In our opinion, these assumptions are physically sound; it is only
their implementation within the mixing entropy argument that is
problematic. We shall avoid that argument by using work 
(more precisely, its maximum in absolute value, ergotropy) 
instead of entropy.

Before mixing, how much work  can be extracted from the total system
containing $M$ separate gases? The answer depends on the specification
of the interaction between the gases and the sources of work. These
interactions are chosen under the following assumptions:

(1)\, Since the gases are ideal, it is natural to assume that the
  sources act on each particle separately, i.e. the sources by
  themselves do not introduce interparticle interactions.

(2)\, Work sources act on the internal degrees of freedom only. This
  is because the internal and the translational degrees of freedom are
  decoupled, and because the translational degrees of freedom are in
  (local) equilibrium, so it is useless to try to extract any work
  from them; recall our discussion around (\ref{thomson}).

(3)\, We allow different sources of work to act on different gases. This
is again reasonable, since the gases start out perfectly separated from
each other. 

Given the above assumptions we are led to the following 
time-dependent, internal Hamiltonian for each gas
\BEA
H^{(\alpha)}(t)=\sum_{i=1}^N \left[\,H^{(i)} 
+ V^{(\alpha,i)}(t)\,\right].
\EEA
Since all particles within the given reservoir are equivalent, we have
\BEA
V^{(\alpha,i)}(t)=V^{(\alpha)}(t),
\EEA
where 
\BEA
V^{(\alpha)}(0)=V^{(\alpha)}(\tau)=0,
\EEA
as required by the cyclic-Hamiltonian feature [compare with (\ref{kura})].

It is now seen that the maximal work 
extractable from the pre-mixed state reads
\BEA
\label{g1}
\W_{\rm i}=\N\sum_{\alpha=1}^M\lambda_\alpha\W(\rho_\alpha,H),
\EEA
where $\W(\rho_\alpha,H)$ defined in (\ref{na}) is the maximal work
extracted from the initial state $\rho_\alpha$ with the initial (and final)
Hamiltonian $H$. Note that $\W_{\rm i}$ is proportional to the total number of
particles $\N=\sum_{\alpha=1}^MN_\alpha$ 
thanks to the above assumptions respecting the ideal gas
structure of the problem.

Let us now determine how much work we can extract after the $M$ gases
have mixed. The above conditions for system-work-source interaction
remain valid except the last one:

(3')\,Since the gases now form a single homogeneous system with the
  density matrix $\rho=\sum_{\alpha=1}^M \lambda_\alpha \rho_\alpha$,
  we cannot enforce the different particles (atoms) to couple to
  different sources of work. At best we can couple the
  ${\cal N}=\sum_{\alpha=1}^MN_\alpha$ particles with the same type of work
  sources. Thus, the physically acceptable cyclic Hamiltonians has the
  form (\ref{kura}).

The resulting maximal work reads from (\ref{mimi}, \ref{na}):
\BEA
\label{g2}
\W_{\rm f}
=\N\,\W\left(\sum_{\alpha=1}^M\lambda_\alpha\rho_\alpha\,,\,H\right).
\EEA

The difference between (\ref{g1}) and (\ref{g2}) 
is defined to be the {\it maximal mixing work} or {\it mixing ergotropy} $\Delta\W$:
\BEA
&&\Delta\W\equiv\W_{\rm i}-\W_{\rm f}\\
\label{mixing_work}
&&=\N\left[\sum_{\alpha=1}^m\lambda_\alpha\W(\rho_\alpha,H)
-\W\left(\sum_{\alpha=1}^M
\lambda_\alpha\rho_\alpha\,,\,H\right)\right],~~~~~~\\
&&=\N
\sum_{k=1}^n\eps_k\left(\,
\p_k-\sum_{\alpha=1}^M\lambda_\alpha\p_{k,\,\alpha}
\,\right),
\label{aa}
\EEA
where we employed (\ref{na}), and where $\p_k$ and $\p_{k,\,\alpha}$
are non-increasingly ordered eigenvalues of $\rho$ and $\rho_\alpha$,
respectively.

The fact that maximal work cannot increase upon mixing, 
\BEA
\label{arzni}
\Delta\W\geq 0,
\EEA
should be obvious from the very construction. Here is, however, the
formal proof. Recall (\ref{nana}) and note that
\BEA
\label{aparan1}
&&{\rm max}_{\,U_\alpha}\,\left(\,{\rm tr}\,[\,H\rho_\alpha\,]-
{\rm tr}\,[\,H\,U_\alpha\,\rho_\alpha\,U^\dagger_\alpha\,]\,\right)\\
=&&{\rm tr}\,[\,H\rho_\alpha\,]-
{\rm tr}\,[\,H\,\widetilde{U}_\alpha\,\rho_\alpha\,
\widetilde{U}^\dagger_\alpha\,]
\\
\label{aparan1.5}
\geq
&&{\rm tr}\,[\,H\rho_\alpha\,]-
{\rm tr}\,[\,H\,U\,\rho_\alpha\,U^\dagger\,],
\label{aparan2}
\EEA
where $\widetilde{U}_\alpha$ is the optimal unitary operator which
maximizes (\ref{aparan1}), and where $U$ is any other unitary
operator, including the one which maximizes 
${\rm tr}\,[\,H\rho-H\,U\,\rho\,U^\dagger\,]$.
The desired (\ref{arzni}) is now recovered via multiplying
(\ref{aparan1}--\ref{aparan2}) by $\lambda_\alpha$ and summing over
$\alpha$.  

The very same argument applies if the maximization in the
definition of $\W$ is carried out over a restricted class of
unitary operators or cyclic Hamiltonians (we assume, of course, 
that this is the same class initially and finally). Analogous to
(\ref{arzni}), we then deduce from (\ref{klm}) that $\Delta\W'\geq 0$.

Turning to the conceptual implications of the mixing work $\Delta \W$,
we note that, of course, $\Delta \W=0$ for $\rho_\alpha=\rho$, when
identical gases are mixed. Moreover, it goes
to zero continuously with $\rho_\alpha\to\rho$. 

\begin{itemize}
\item
We therefore consider this continuity of maximally extractable work
as the resolution of the Gibbs paradox within quantum thermodynamics.
\end{itemize}

The first objection for the entropic argument|see our discussion
around (\ref{cl}) and Ref.~\cite{DD}|is now harmless, since now the
concept of thermodynamical reversibility is not employed anywhere; the
machinery of the maximal work-extraction is based completely on 
quantum mechanics alone. As we stressed repeatedly, work is a 
first-principle concept, more fundamental than entropy \cite{Minima}
\footnote{In phenomenological thermodynamics, the problem of the
  maximal work extraction is treated by employing the reversibility
  concept and features of entropy \cite{Landau}. In our opinion, this
  is the reason why the concept of work|though mentioned as a helpful
  one for interpreting the Gibbs paradox \cite{Lande,BG}|was never
  seriously employed for resolving the paradox.}.

Note that when $\rho_1$ and $\rho_2$ are pure states, the converse of
the above statement appears to be valid: if $\rho_1$ and $\rho_2$ are
different, then $\Delta\W >0$. This is because the only pure state
that cannot provide work is the ground state of the Hamiltonian
$H$. If, however, at least one of the two density matrices is mixed,
there are different states $\rho_1$ and $\rho_2$ such that
$\Delta\W=0$. For the simplest example recall (\ref{thomson}) and take
as $\rho_1$ and $\rho_2$ two equilibrium states with different
temperatures $T_1$ and $T_2$.

To illustrate the above statements in more detail, we
turn to the density matrices given by 
(\ref{kaban}--\ref{kaban2}), where the Hamiltonian
$H$ has two energy levels $0$ and $\eps>0$.  
Recalling (\ref{kabo}) we
get from (\ref{mixing_work}) that $\Delta\W$ 
is a simple function of the overlap:
\BEA
\Delta\W=\frac{\N\eps}{2}\left[\,
1-|\langle a_1|a_2\rangle|\,
\right].
\label{wit}
\EEA For completely distinguishable, classical states $|\langle
a_1|a_2\rangle|=0$ this gives $\Delta \W=\eps/2$, while for identical
states $|\langle a_1|a_2\rangle|=1$, $\Delta \W=0$. The classical
argument describes {\it only} these extremes 
(i.e., completely different or identical) and, thus, creates the
paradox.

\section{
How the mixing work depends on the available instruments.}
\label{instruments}

Let us now turn to the second objection against the entropic argument.
We recall from section \ref{o} that once the difference between two
states is recognized to be an operational notion|two states may not
differ under inspection by some instruments, but turn out to be
different if more refined ones are used|we should expect that this feature
is reflected in a satisfactory resolution of the Gibbs paradox.

As we stressed repeatedly, the notion of available work is operational
in the above sense. So is the mixing work defined in
(\ref{mixing_work}). Moreover, the situation is non-trivial, since
$\Delta\W$ can both increase or decrease under restricting the
available instruments (i.e. system-work-source interactions), as we
show now.

To illustrate this fact, let us take the internal Hilbert space of
all particles having two dimensions (e.g., spin-$\frac{1}{2}$):
\BEA
\label{kozma}
\rho_\alpha=\half\left(1+\vec{n}_\alpha\,\vec{\sigma}\right),\qquad
\alpha=1,...,M,
\EEA
where $\vec{\sigma}=\left(\, \sigma_1,\sigma_2,\sigma_3\,\right)$ 
are the Pauli ($2\times 2$) matrices, and where
\BEA
\vec{n}_\alpha=(n_{1,\,\alpha},n_{2,\,\alpha},n_{3,\,\alpha}),\qquad
|\vec{n}_\alpha|\leq 1
\EEA
is the Bloch c-vector. Recalling the spectrum
\BEA
{\rm Spec}\,\{\,\rho_\alpha\,\}=\frac{1}{2}\left(
1\pm |\vec{n}_\alpha|
\right),
\EEA
we get from (\ref{mixing_work}),
\BEA
\label{bet}
\Delta\W=\frac{\N\eps}{2}\left(\,
\sum_{\alpha=1}^M\lambda_\alpha |\vec{n}_\alpha|-\left|
\sum_{\alpha=1}^M\lambda_\alpha \vec{n}_\alpha \right|\,
\right).
\EEA

On the other hand, if for the Hamiltonian
\BEA
H=\frac{\eps(1+\sigma_3)}{2}, 
\EEA
the maximization over the uninary operators
in (\ref{na}) is carried out only over those unitary operators 
which permute the diagonal
elements of the corresponding density matrices in the energy
representation [compare with (\ref{klm}, \ref{klm1})], the mixing work
will read
\BEA
\label{shtani}
\Delta\W'=\frac{\N\eps}{2}\left(\,
\sum_{\alpha=1}^M\lambda_\alpha |n_{3,\,\alpha}|-\left|
\sum_{\alpha=1}^M\lambda_\alpha  n_{3,\,\alpha} \right|\,
\right),
\EEA
where $n_{3,\,\alpha}$ is the third component of the vector
$\vec{n}$.

It is obvious that there are cases where
\BEA
\Delta\W>
\Delta\W',
\EEA
e.g., choose $n_{3,\,\alpha}$ all having the same sign which leads to
$\Delta\W'=0$. It is,
however, less expected that there can also be situations where
\BEA
\label{kust}
\Delta\W'>
\Delta\W.
\EEA
This means:
\begin{itemize}
\item
use of less precise instruments can increase the amount of mixing work.
\end{itemize}

To show this, let us choose the case
\BEA
\label{swi}
\sum_{\alpha=1}^M 
\lambda_{\alpha} n_{3,\,\alpha}=0,
\EEA
and write from (\ref{bet}, \ref{shtani})
\BEA
&&\frac{\Delta\W'-\Delta\W}{\N\,(\varepsilon/2)}~~~~~~~~~~~~~~~~~~~~~~~~~~~~~
\nonumber\\
\label{choban}
&&=\sum_{\alpha=1}^M \left(
\sqrt{\lambda^2_\alpha  n^2_{3,\,\alpha}  }-
\sqrt{\lambda^2_\alpha\,\left[  n^2_{3,\,\alpha} 
+n^2_{1,\,\alpha}+n^2_{2,\,\alpha} \right]}\,
\right)~~~~~~~\\
&&+\sqrt{\left( \sum_{\alpha=1}^M \lambda_\alpha  n_{1,\,\alpha} \right)^2
+\left(\sum_{\alpha=1}^M \lambda_\alpha  n_{2,\,\alpha} \right)^2
}.
\EEA

In (\ref{choban}) we use the inequality\footnote{To prove (\ref{bazar}) 
make an incomplete Taylor expansion for $f(x)=\sqrt{x}$:
$f(x+y)=f(x)+yf'(x) +\frac{y^2}{2}
f''(\xi) $, where $x\leq \xi\leq x+y$,
and disregard $\frac{y^2}{2} f''(\xi)\leq 0$.}
\BEA
\label{bazar}
\sqrt{x}-\sqrt{x+y}\geq -\frac{y}{2\sqrt{x}}.
\EEA

Taking for simplicity $\lambda_\alpha  n_{1,\,\alpha}=\lambda_\alpha
n_{2,\,\alpha}= b$ and $|\lambda_\alpha  n_{3,\,\alpha}|=a$|and thus  
$M$ should be even to satisfy (\ref{swi})|we get
\BEA
\Delta\W'-\Delta\W
\geq \sqrt{2}\, M\,\N\,|b|\left(
1-\frac{|b|}{a\sqrt{2}}
\right).
\label{qq}
\EEA
By suitable choice of $a$ and $b$, one can make the RHS of
(\ref{qq}) positive, thus proving the desired statement (\ref{kust}).

\section{
Mixing work and the degree of mixing.}
\label{degree}

As we saw, the mixing work is zero when there is no true mixing,
i.e., when the internal states of the mixed gases are identical
\footnote{\label{madras}Note that when the overall numbers of particles
  $N_\alpha$ in each reservior is not very large, even the mixing of
  completely identical gases brings about changes in their final state
  \cite{GLP}. This is due to different fluctuation characteristics of
  the translational motion \cite{GLP}, e.g., before mixing the number
  of particles in the volume $V_\alpha$ is precisely $N_\alpha$, while
  after mixing this number of particles will fluctuate being equal to
  $N_\alpha$ only on average.  We shall neglect this effect assuming
  $N_\alpha$ to be sufficiently large.}. It is expected that the
  mixing work will decrease together with the degree of mixing.

Consider the mixing work $\Delta \W (\vec{\lambda})$ as a function of
the weights $\vec{\lambda}=\{\lambda_\alpha \}_{\alpha=1}^M$.  For
fixed states $\{\rho_\alpha \}_{\alpha=1}^M$, we expect that
if $\vec{\lambda}$ is more inhomogeneous than $\vec{\mu}$, then 
\BEA
\Delta\W(\vec{\mu})
\geq
\Delta\W(\vec{\lambda}).
\label{comrad}
\EEA
Here is an exaggerated example illustrating (\ref{comrad}): for two
species the degree of mixing is expected to be higher when having 100
particles of each type than when having 199 and 1, respectively.
The weights for this example are, respectively, $\mu_1=\mu_2=\frac{1}{2}$
and $\lambda_1=\frac{199}{200}$, $\lambda_2=\frac{1}{200}$.

Below we clarify in which sense the intuitive expectation (\ref{comrad}) is
correct.

\subsection{Majorization.}

First we need the proper formalization for the notion of
``inhomogeneous''. This is provided by the concept of majorization
\cite{major} which we shortly recall below.

For two sets of probabilities $\vec{\lambda}= \{\lambda_\alpha
\}_{\alpha=1}^M$ and $\vec{\mu}=\{\mu_\alpha \}_{\alpha=1}^M$,
$\vec{\lambda}$ majorizes $\vec{\mu}$ (i.e., $\vec{\lambda}$ is
more inhomogeneous than $\vec{\mu}$),
denoted as 
\BEA
\vec{\mu}\maj \vec{\lambda}, 
\EEA
if for all $1\leq m\leq M$
\BEA
\label{durs}
\sum_{\alpha=1}^m\lambda^{\downarrow}_\alpha
\geq \sum_{\alpha=1}^m \mu^{\downarrow}_\alpha,
\EEA
where $\vec{\lambda}^{\downarrow}$ means non-increasing ordering 
of $\vec{\lambda}$ [recall (\ref{shut})].

To illustrate (\ref{durs}): the uniform vector $(1/M,...,1/M)$ is
majorized by all other probability vectors, while any deterministic
vector, e.g.  $(1,0..,0)$, majorizes all others.  It follows from
(\ref{durs}) that $\sum_{\alpha=1}^M f(\lambda_\alpha)\leq
\sum_{\alpha=1}^M f(\mu_\alpha)$ for any concave function $f(x)$ \cite{major},
e.g., $f(x)=-x\ln x$ (entropy).
  
  The majorization property is transitive: $\vec{\mu}\maj
  \vec{\lambda}$ and $\vec{\lambda}\maj \vec{\nu}$ imply $\vec{\mu}\maj
  \vec{\nu}$. Also $\vec{\lambda}\maj \vec{\mu}$ and $\vec{\mu}\maj
  \vec{\lambda}$, imply $\vec{\lambda^{\downarrow}}=
  \vec{\mu^{\downarrow}}$.  However, this property is incomplete: for
  $n\geq 3$ there are vectors $\vec{\lambda}$ and $\vec{\mu}$ for
  which neither $\vec{\lambda}$ majorizes $\vec{\mu}$, nor does $\vec{\mu}$
  majorize $\vec{\lambda}$ \cite{major}.

\subsection{Quasi-classical situation.}

Let the initial states of the gases be $M$ pure, orthonormal states
\BEA \rho_\alpha= |\psi_\alpha\rangle\langle\psi_\alpha|, \qquad
\langle\psi_\alpha|\psi_\beta\rangle =\delta_{\alpha\beta}.
\EEA

We call this situation quasi-classical, since following the original
formulation of the Gibbs paradox within classical thermodynamics,
the internal states are completely distinguishable and provide
definite values for any observable that has
$\{|\psi_\alpha\rangle\}_{\alpha=1}^M$ as its eigenfunctions.

Let us now prove that
if $\vec{\lambda}$ is more inhomogeneous than $\vec{\mu}$, 
i.e., if (\ref{durs}) holds, then inequality (\ref{comrad}) is valid.
To this end
we first employ summation by parts
\BEA
\sum_{k=1}^n\eps_k\lambda_k^\downarrow
=&&
\eps_n-(\eps_2-\eps_1)\lambda_1^\downarrow
-(\eps_3-\eps_2)(\lambda_1^\downarrow+\lambda_2^\downarrow)\nonumber\\
-&&(\eps_4-\eps_3)(\lambda_1^\downarrow+\lambda_2^\downarrow
+\lambda_3^\downarrow)
-...\nonumber,
\EEA
and then recalling (\ref{aa}) we get 
\BEA
&&\frac{\Delta\W(\vec{\mu})-\Delta\W(\vec{\lambda})}{\N}=
\sum_{k=1}^n\eps_k\left[\mu_k^\downarrow
-\lambda_k^\downarrow
\right]\nonumber\\
&&=(\eps_2-\eps_1)(\lambda_1^\downarrow-\mu_1^\downarrow)
+(\eps_3-\eps_2)(\lambda_1^\downarrow+\lambda_2^\downarrow
-\mu_1^\downarrow-\mu_2^\downarrow)\nonumber\\
&&+...\geq 0.
\EEA
Here each separate term is non-negative due to (\ref{hu}, \ref{durs}).

\begin{itemize}
\item
For this quasi-classical situation the above intuition (more
mixing means larger mixing work) is correct.
\end{itemize}

\subsection{Quantum situation.}

Let us assume that the initial states $\{\rho_\alpha\}_{\alpha=1}^M$
are not orthogonal. For simplicity we shall work with the simplest
non-trivial situation: 
\BEA
n=M=2,
\EEA
i.e., two-dimensional internal state and
two mixed gases. $\Delta\W$ is now given by (\ref{bet}).
We assume that $\vec{\lambda}$ is more ordered than $\vec{\mu}$ in the
sense of majorization, which for $M=2$ implies:
\BEA
\label{castro}
\lambda_1\geq\lambda_2,
\qquad
\mu_1\geq\mu_2,
\qquad
\lambda_1\geq\mu_1.
\EEA
Note that for the considered two-dimensional situation, $n=2$, 
the majorization order coincides, e.g., with the entropic order:
Eq.~(\ref{castro}) implies 
$-\lambda_1\ln\lambda_1-\lambda_2\ln\lambda_2
\leq -\mu_1\ln\mu_1-\mu_2\ln\mu_2 $.

We now intend to clarify under which 
conditions the inequality (\ref{comrad}) holds. 
Recalling (\ref{bet}) this inequality is equivalent to
\BEA
\label{samson}
&&(\,\mu_1-\lambda_1\,)\left( \,|\vec{n}_1| -|\vec{n}_2|
  \,\right)
\\
\geq 
&&\sqrt{\,\left(\,
\mu_1|\vec{n}_1|+\mu_2|\vec{n}_2|\,\right)^2 
-2\mu_1\mu_2|\vec{n}_1|\,|\vec{n}_2|(1-\cos\phi) 
\,}
\nonumber\\
-&&\sqrt{\,\left(\,
\lambda_1|\vec{n}_1|
+\lambda_2|\vec{n}_2|\,\right)^2 -2\lambda_1\lambda_2
|\vec{n}_1|\,|\vec{n}_2|(1-\cos\phi) 
\,},
\nonumber
\EEA
where $\cos\phi$ is defined as
\BEA
\label{ja}
\vec{n}_1\cdot\vec{n}_2
=|\vec{n}_1|\,|\vec{n}_2|\,\cos\phi.
\EEA
When both states are pure, $|\vec{n}_1|=|\vec{n}_2|=1$, inequality
(\ref{samson}) reduces to $\mu_1\mu_2\geq \lambda_1\lambda_2$ or
\BEA
\label{bek}
\lambda_1+\mu_1\geq 1,
\EEA
a condition which is always satisfied in view of (\ref{castro}).

Assume in (\ref{samson}) that $\phi$ is small, and expand
(\ref{samson}) to first order of $1-\cos\phi$. After algebraic steps
we get a generalization of (\ref{bek})
\BEA
\label{gadi}
\lambda_1+\mu_1
\geq 1+\lambda_1\mu_1
\left[\,1-\frac{|\vec{n}_1|}{|\vec{n}_2|}\,\right].
\EEA
This inequality is already not always satisfied. When
$|\vec{n}_1|/|\vec{n}_2|$ is sufficiently small, 
i.e., one of the states is considerably more mixed,
Eq.~(\ref{gadi}) may be violated;
take, e.g., $\lambda_1=0.8$ and $\mu_1=0.7$. We conclude that

\begin{itemize}

\item In the quantum situation the mixing work may be a non-monotonous
  function of the degree of mixing, though it goes to zero continuously 
  when the substances become identical.

\end{itemize}

\section{Distinguishability and Mixing.}
\label{dis}

Another way to control the mixing is to keep the weights equal, but
make the internal states $\rho_1$ and $\rho_2$ closer to each
other. It is natural to ask whether the mixing
work is a monotonic function of the difference between these
substances, i.e., whether decreasing this difference always makes the
mixing work smaller. Below we are going to show that this is not
always the case, though the mixing work, of course, goes to zero 
in the limit of identical substances.

First of all we need a clear understanding of the proper distance
(closeness) between two density matrices $\rho_1$ and $\rho_2$. 
The answer is trivial for pure states as in (\ref{kaban}): 
any monotonic function of the overlap
\BEA
\label{kkk}
{\rm tr}\,(\rho_1\rho_2)=
|\langle a_1|a_2\rangle|^2
\EEA 
can be taken as the proper degree of closeness.

The generalization of the overlap (\ref{kkk}) to mixed states is also
well known and was derived from several different perspectives
\cite{Bures,Caves}. This ``distinguishability'' reads:
\BEA
d(\rho_1,\rho_2)=\left[\,{\rm
tr}\,\left(\sqrt{\rho_1^{1/2}\rho_2\,\rho_1^{1/2}}
\,\right)\right]^2.
\EEA
Let us note that $d(\rho_1,\rho_2)$ is symmetric 
\BEA
d(\rho_1,\rho_2)=d(\rho_2,\rho_1), 
\EEA
concave
\BEA
d(\rho, x\rho_1+(1-x)\rho_2)\geq xd(\rho,\rho_1)+(1-x)d(\rho,\rho_2), 
\EEA
and varies between $0$ and $1$, 
\BEA
0\leq d(\rho_1,\rho_2)\leq 1,
\EEA
being equal to $1$ if and only if $\rho_1=\rho_2$. It is also
multiplicative
\BEA
d(\rho_1\otimes\rho_3,\rho_2\otimes\rho_4)
=d(\rho_2,\rho_1)d(\rho_3,\rho_4), 
\EEA
invariant under unitary transformations,
\BEA
d(\rho_1,\rho_2)=d\left(U\,\rho_1\,U^\dagger\,,\,U\,\rho_2\,U^\dagger\right),
\qquad U^\dagger U=1,
\EEA
it increases under completely positive evolution,
and reduces to ${\rm tr}\, (\, \rho_1\rho_2    \,)$
if $\rho_1$ or $\rho_2$ is pure.

In particular, $d(\rho_1,\rho_2)$ has the proper information-theoretic
meaning as arising from the statistical distance between the data
acquired by optimal measurements carried out for distinguishing between
$\rho_1$ and $\rho_2$ \cite{Caves}.

In Appendix \ref{molod} we determine $d(\rho_1,\rho_2)$ for two spin $\half$
density matrices $\rho_1$ and $\rho_2$, given as in (\ref{kozma}), with
Bloch vectors $\vec{n}_1$ and $\vec{n}_2$, respectively:
\footnote{Note the difference with $2{\rm tr}\,(\rho_1\rho_2)-1=
\vec{n}_1\cdot\vec{n}_2$.}
\BEA
2d(\rho_1,\rho_2)-1=\vec{n}_1\cdot\vec{n}_2
+\sqrt{1-|\vec{n}_1|^2}\,
\sqrt{1-|\vec{n}_2|^2}.
\label{texas}
\EEA
For pure states $|\vec{n}_\alpha|=1$, and we expectedly obtain from
(\ref{texas}) propotionality between the overlap and the scalar
product of the two Bloch vectors.

For the mixing work we have from (\ref{bet})
\BEA
\label{bet1}
&&\frac{\Delta\W}{\N}=\frac{\eps}{2}\left(\,
\lambda_1|\vec{n}_1|+\lambda_2|\vec{n}_2|\right.\nonumber\\
&&-\left. \sqrt{\lambda_1^2|\vec{n}_1|^2+\lambda_2^2|\vec{n}_2|^2
+2\lambda_1\lambda_2\,\,\vec{n}_1 \cdot \vec{n}_2\,
}
\right).
\EEA

When comparing (\ref{bet1}) with (\ref{texas}) we see that if only the
scalar product $\,\vec{n}_1 \cdot \vec{n}_2\,$ is varied|with the
modules $|\vec{n}_1|$ and $|\vec{n}_2|$ being fixed|making the two
states closer, the mixing work $\Delta\W$ indeed monotonically
decreases. In particular, this is the case for pure states $\rho_1$
and $\rho_2$.  However, as seen from (\ref{texas}),
for mixed states $\rho_1$ and $\rho_2$ the
scalar product between the corresponding Bloch vectors 
is only one aspect of closeness. To look at another setup, vary $|\vec{n}_1|$
with $|\vec{n}_2|$ while keeping their mutual angle $\phi$ fixed (see (\ref{ja})
for the definition of $\phi$). 
Note that $\Delta \W$ always increases with $|\vec{n}_1|$:
\BEA
&&\frac{\partial \Delta\W}{ \partial |\vec{n}_1| }=
\frac{\eps}{2\,|\lambda_1\vec{n}_1+\lambda_2\vec{n}_2|
}\times\nonumber\\
&&\left(
|\lambda_1\vec{n}_1+\lambda_2\vec{n}_2|
-\lambda_1|\vec{n}_1|-\lambda_2|\vec{n}_2|\,\cos\phi
\right)\geq 0.
\label{babakhan}
\EEA
On the other hand, we have from (\ref{texas})
\BEA
&&\frac{\partial d(\rho_1,\rho_2) }{ \partial |\vec{n}_1| }=
\frac{1}{2\,\sqrt{1-|\vec{n}_1|^2}\,}\times\nonumber\\
&&\left(\,
\cos\phi\, |\vec{n}_2|\,
\sqrt{1-|\vec{n}_1|^2}\,
-\, |\vec{n}_1|\,
\sqrt{1-|\vec{n}_2|^2}
\,\right)
\label{baa}
\EEA
When the scalar product is positive: $\,\vec{n}_1 \cdot
\vec{n}_2\,=\cos\phi>0$, Eq.~(\ref{baa}) can be positive, i.e., 
the states can get closer with increasing $|\vec{n}_1|$, if
$|\vec{n}_1|$ is sufficiently small, or if $|\vec{n}_2|$ is
sufficiently close to $1$.
Comparing with (\ref{babakhan}) we conclude:

\begin{itemize}
\item
It is possible to make the two states of the mixing substances
closer to each other and simultaneously increase the mixing work
\footnote{Note that Ref.~\cite{Shu} discusses a similar situation in
classical chemical physics. The analogy, however, appears to be
superficial, since the author of Ref.~\cite{Shu} bases his conclusions
on the non-additive classical formula $S_{\rm cl}(N,V)=N\ln{V}$ for
entropy.  }.

\end{itemize}

We stress that all conclusions of the present section are valid under other
reasonable measures of distance between $\rho_1$ and $\rho_2$, e.g.,
${\rm tr}\,\left[ (\rho_1-\rho_2)^2 \right]$. Indeed, it amounts to a
simple check that the qualitative conclusion we got after (\ref{baa}) is
valid as well for this measure of closeness. 

\section{Conclusion.}

Since its formulation in the late 1870's, the Gibbs paradox has, lacking a simple solution, 
become a quest for the understanding of phenomenological thermodynamics from a more fundamental 
theory. This attempt to go to a deeper level is the reason for its 
importance \cite{S,T,Lande,Klein,LP,GLP}. 
Its understanding happens to have several layers. First, it was realized
that it is necessary to take into account explicitly  the difference
between the particles, which drives the classical formulation of the
paradox, but how much they differ shows up nowhere in formulas. 
Together with the separation of characteristic relaxation times and the von Neumann
definition of entropy, this brought about the quantum mixing entropy
argument which for many years was seen as the resolution of the
Gibbs paradox
\cite{Lande,Klein,LP,GLP,Lesk}. It was, however, pointed out that the
argument introduces a new conceptual difficulty precisely when it
claims to solve the paradox \cite{DD}. The details being presented in
section \ref{argument}, we simply recall that this difficulty has to
do with the features of entropy, more precisely, with the fact that
the entropy is not a sufficiently primitive (first-order) quantity
in the  situation at hand.
So a deeper reduction level has to be involved for the resolution of
the Gibbs paradox.

In our opinion, the basic reason why classical thermodynamics fails for the
understanding of mixing entropy is that the difference between an A atom
and a B atom is not dealt with properly, in particular, because no
macroscopic limit is involved in differences between gases A and B. Lacking
such a limit, the basis for phenomenological thermodynamics, be it based
on classical or quantum statistical physics, has disappeared and its
application indeed leads to paradoxes and ill-defined issues such as the
non-operational nature of the mixing entropy.  We are thus left with the
search for a more fundamental approach. Such a possibility is offered 
by the field of quantum thermodynamics, that has been considered in 
recent years by several groups, see e.g. 
\cite{ABN,ScullyDemon,Dom_power,ANbkj,Minima,NA,Landauer,Kosloff,Mahler,
Scully,ABNQthermo,Ruben,Dom_frame,Dom_Wo}.
 
In the current paper we have presented an explanation of the Gibbs
paradox within this field. Here the notion of entropy is known to be
easily blurrred, and a paradigm shift is called for towards the more
ancient concept of work [energy transferred to macroscopic work
sources], which still plays a clear and empirical role. In particular,
quantum thermodynamics applies to finite systems, e.g., the basic
formulations of the second law are well-defined both conceptually and
operationally \cite{Minima}.  Indeed, it could be shown that Landauer's
principle that connects the minimal energy dispersion to erase one bit
of information $\Delta Q\ge kT\ln 2$ may loose its validity in the
domain of quantum thermodynamics~\cite{Landauer}, while the Maxwell
demon problem just found new viewpoints there ~\cite{ANdemon,
ScullyDemon}.  

It was further shown that the maximally extractable work
(which we called `ergotropy' in an earlier paper with R. Balian
\cite{ABN}) can be clearly defined before and after the mixing process.
The difference between them defines the maximal mixing work, or mixing ergotropy,
a non-negative quantity which smoothly goes to zero when the substances become more and more
equal to each other, as it is for a single substance, thus solving the Gibbs paradox
in the work formulation.  (As should be clear from our presentation,
we consider that the Gibbs paradox in its entropic formulation 
has not been properly solved so far, and that we even do not believe 
that it is consistently resolvable in that form.)

In contrast to entropy, the features of work can be directly based on
the first principles of quantum mechanics and are well-defined for any
(equilibrium or non-equilibrium) state of a system interacting with
macroscopic sources of work. In particular, there is no need to involve
features of thermodynamical reversibility for defining and interpreting
the mixing work; see in this context Footnotes \ref{conrad1} and \ref{conrad2}. 
On top of that, the concept of maximal work has a
well-defined operational character, because it is always defined with
respect to a definite class of work-sources acting on the system of
interest.  The features of work and entropy are recalled and contrasted
in sections \ref{wowork}, \ref{contraentropy} and Appendix
\ref{takhtak}.  Recall in this context that the concept of work was already
employed in the literature devoted to the Gibbs paradox \cite{Lande,BG},
but its potential applications were conceived in the framework of
phenomenological thermodynamics. In that way, they encounter almost all
objections raised against the mixing entropy argument.  Only after the
problem of maximal work-extraction was solved from the first principles
\cite{ABN}, it became possible to approach an explanation of the
Gibbs paradox with the help of the mixing work. This explanation is free of
the difficulties which plagued the quantum entropy argument. 

To keep our approach as natural as possible, we have supposed that,
after allowing the gases to mix,  the translational degrees of freedom 
equilibrate rather quickly, while their spin degrees of freedom
do not equilibrate at all at the timescales for which our discussion applies
because their dynamics are supposed to 
take place on a much larger time scale. 
For this reason, these degrees of freedom can be considered
as not coupled to the bath, which saves us from discussing the more
complicated situation where heat exchange of the spins with the bath 
would also matter.

The consistent resolution of the paradox presents features that might not
have been anticipated before. It appears that less precise control can,
depending on the situation, bring a larger or a smaller amount of mixing work. 
We have also seen that a naive intuition relating the degree of mixing and
the distinguishability with the mixing ergotropy may not always be correct:
sometimes making the initial states of the mixed substances closer to
each other (in the proper information-theoretic sense) can make the
amount of mixing work larger. These are warnings against a direct
association of physical irreversibility (i.e., mixing work) with 
lack of information: while the amount of mixing work is non-zero due to less
information on the identity of atoms in the post-mixed state, the
relation of this lack of information to the physical irreversibility
can be non-trivial and counter-intuitive. 

\acknowledgments It is a pleasure to thank Roger Balian for inspiring
discussions and Peter Keefe for carefully reading the proofs. 

A.E. A. acknowledges hospitality at the University of Amsterdam. His
work was partially supported by the Stichting voor Fundamenteel Onderzoek der
Materie (FOM, financially supported by the Nederlandse Organisatie
voor Wetenschappelijk Onderzoek (NWO)) and by CRDF grant ARP2-2647-YE-05.

\section*{
APPENDIX A: THE FORMULATION OF THE SECOND LAW SETS THE CHOICE OF THE ENTROPY}

\label{takhtak}
Here we shall recall why entropy is not uniquely defined and why its
possible definitions depend on the second law of thermodynamics.

Consider an adiabatically isolated process done on a quantum
system described by density matrix $\rho(t)$.  The process is realized
via a time-dependent Hamiltonian, with the cyclic feature defined
according to Eq.~(\ref{cycle}). The evolution of the system starts from
some Gibbsian equilibrium state at a positive temperature.  It is well
known from thermodynamics \cite{Landau,Balian} and can be derived from
the first principles of quantum mechanics|see \cite{Lenard,Lindblad} and
our discussion in section \ref{maxi}|that in this process the system
consumes positive work which is the statement of the second law in
Thomson's formulation. It is natural to look for the counterpart of this
formulation in terms of entropy. Since the dynamics of the thermally
isolated system is unitary, the von Neumann entropy $S_{\rm
vN}(\rho(t))$ is constant in time; so it is not suitable for being the
counterpart of the Thomson's formulation. This argument is sometimes
dismissed on the ground that the unitary dynamics is reversible and thus
the constant behavior of the von Neumann entropy is reasonable.  In the
present context this seems incorrect, in particular, because a positive
amount of work is put into the system in accordance with Thomson's
formulation of the second law. 

In the spirit of the relevant entropy approach \cite{Balian} (there are
many entropies each one for its own situation and its own use) we can
regard as physical another entropy 
\BEA S_{\rm T}(t)=-\sum_k \pi_k(t)\ln \pi_k(t), \EEA 
with $\pi_k$ being the time-dependent probabilities of
various values of the system's energy (given by the time-dependent
Hamiltonian) in the state $\rho(t)$. This definition of entropy was
proposed and advocated by Tolman \cite{Tolman}.  For the considered
process, $S_{\rm T}$ does have several reasonable properties:

(1) At the end of the cyclic-Hamiltonian process $S_{\rm T}$ is larger than in the
initial equilibrium state \cite{Lenard,Lindblad}.  

(2) Under conditions specified in \cite{Minima}, the change of $S_{\rm T}$
is minimal for the adiabatically slow process, again as required by
thermodynamics. 

(3) $S_{\rm T}$ is maximal in equilibrium. 

Each of these three features corresponds to a specific formulation
of the second law. The features (1) and (2) will not be valid when using
the von Neumann entropy.

Thus, we are led to employ the Tolman definition of entropy following to
the requirements of the second law.

Let us now consider an isothermal
process, where the system (e.g., a spin or a brownian particle) weakly
interacts with an equilibrium thermal bath at temperature $T$.
The bath being in equilibrium means for the present context
two things. First, it starts in the equilibrium state at temperature $T$,
and, second, its relevant characteristic times are much larger than those
of the system. (An additional feature of weak interactions was stressed
by us above.) It is again well known from phenomenological
thermodynamics, and is derived from the first principles of quantum
mechanics that during the relaxation of the system to equilibrium, the
(non-equilibrium) free energy decays, a statement known as H-theorem
\cite{Landau,Balian,Lindblad}:
\BEA
\label{H}
&&\frac{\d F}{\d t}
=\frac{\d }{\d t}\left[E(t)-TS_{\rm vN}(t)\right]
\nonumber\\
&&\equiv
\frac{\d }{\d t}\left[
{\rm tr}\,(\rho(t)\, H)+T
{\rm tr}\,(\rho(t)\, \ln\rho(t)\,)
\right]\leq 0,
\EEA 
where $\rho(t)$ is the density matrix of the system, and where $H$ is
its time-independent Hamiltonian. Note especially that the H-theorem
will in general not be valid if instead of the von Neumann entropy we
shall use in (\ref{H}) the Tolman entropy $S_{\rm T}$. Thus, here for
isothermal processes we had to return to the von Neumann definition of
entropy. What is the proper definition of entropy when the process is
neither isothermal nor thermally isolated is in general not known
\cite{NA}. 

In short, in statistical physics the definitions of entropy are
contextual, since they already depend on various formulations of the
second law.  It is, therefore, questionable whether arguments based on
entropies are able to resolve thermodynamical paradoxes.

\section*{ APPENDIX B: A USEFUL IDENTITY}
\label{molod}

Here we outline how to calculate the overlap defined in (\ref{texas})
for two spin $\half$ density matrices
\BEA
\rho=\half\left(1+\vec{n}_\alpha\cdot\vec{\sigma}\right),\qquad \alpha=1,2.
\EEA

We need the following facts. First, note that the square root of $\rho$ is most
conveniently calculated when representing $\rho$ as
\BEA
\rho=\half\left(1+\sin\theta\,\vec{\chi}\cdot\vec{\sigma}\right),
\EEA
where $0\leq \theta\leq \pi/2$, and where $\vec{\chi}$ is a unit
vector $|\vec{\chi}|=1$. Then
\BEA
\sqrt{\rho}=
\sqrt{\half}
\left(\cos\frac{\theta}{2}
+\sin\frac{\theta}{2}\,\vec{\chi}\cdot\vec{\sigma}\right),
\EEA
and
\BEA
{\rm tr}\,\sqrt{\rho}=\sqrt{2}\,\cos\frac{\theta}{2}
=\sqrt{1+\sqrt{1-|\vec{n}|^2}}.
\EEA
Next, we need the known identity for Pauli matrices
\BEA
\label{barbos}
\left(\,\vec{n}_1\cdot\vec{\sigma}\,\right)
\,
\left(\,\vec{n}_2\cdot\vec{\sigma}\,\right)
=(\vec{n}_1\cdot\vec{n}_2)
+i\,\vec{\sigma}\cdot
[\,\vec{n}_1\times\vec{n}_2\,],
\EEA
where $[\,\vec{n}_1\times\vec{n}_2\,]$ is the vector product.
And, finally, the last ingredient is given by
\BEA
\label{rt}
\left(\,\vec{\chi}\cdot\vec{\sigma}\,\right)
\,
\left(\,\vec{\xi}\cdot\vec{\sigma}\,\right)\,
\left(\,\vec{\chi}\cdot\vec{\sigma}\,\right)
=\left(\,2\,(\vec{\chi}\cdot\vec{\xi})\,\vec{\chi}
-\vec{\xi}\,\right)\cdot\vec{\sigma},
\EEA
where $\chi$ and $\xi$ are unit vectors. Eq.~(\ref{rt})
follows from (\ref{barbos})
and the double vector product
identity:
\BEA
[\,\vec{n}_3\times
[\,
\vec{n}_1\times
\vec{n}_2\,]\,]=
\vec{n}_1\,(\vec{n}_2\cdot\vec{n}_3)
-\vec{n}_2\,(\vec{n}_1\cdot\vec{n}_3).
\EEA

\end{document}